\documentclass[8.5pt,twoside,twocolumn]{article}
\PassOptionsToPackage{super,sort&compress,comma}{natbib} 
\usepackage[version=3]{mhchem}
\usepackage{times,mathptmx}
\usepackage{sectsty}
\usepackage{balance} 

\usepackage{graphicx} 
\usepackage{lastpage}
\usepackage[format=plain,justification=raggedright,singlelinecheck=false,font=small,labelfont=bf,labelsep=space]{caption} 
\usepackage{fancyhdr}
\usepackage{textcomp}
\usepackage{xspace}

\usepackage[usetitle=true,usedoi=true,linkdoi=true,mciteplus=false]{rsc}
\usepackage{hyperref}

\newcommand{\Tc}{\ensuremath{T_\mathrm{c}}}
\newcommand{\kB}{\ensuremath{k_\mathrm{B}}}

\newcommand{\ea}{\textit{et al.}\xspace}

\begin{document}

\thispagestyle{plain}
\fancypagestyle{plain}{
\renewcommand{\headrulewidth}{1pt}}
\renewcommand{\thefootnote}{\fnsymbol{footnote}}
\renewcommand\footnoterule{\vspace*{1pt}%
\hrule width 3.4in height 0.4pt \vspace*{5pt}} 
\setcounter{secnumdepth}{5}

\makeatletter 
\def\subsubsection{\@startsection{subsubsection}{3}{10pt}{-1.25ex plus -1ex minus -.1ex}{0ex plus 0ex}{\normalsize\bf}} 
\def\paragraph{\@startsection{paragraph}{4}{10pt}{-1.25ex plus -1ex minus -.1ex}{0ex plus 0ex}{\normalsize\textit}} 
\renewcommand\@biblabel[1]{#1}            
\renewcommand\@makefntext[1]%
{\noindent\makebox[0pt][r]{\@thefnmark\,}#1}
\makeatother 
\renewcommand{\figurename}{\small{Fig.}~}
\sectionfont{\large}
\subsectionfont{\normalsize} 

\fancyfoot{}
\fancyfoot[RO]{\footnotesize{\sffamily{1--\pageref{LastPage} ~\textbar  \hspace{2pt}\thepage}}}
\fancyfoot[LE]{\footnotesize{\sffamily{\thepage~\textbar\hspace{3.45cm} 1--\pageref{LastPage}}}}
\fancyhead{}
\renewcommand{\headrulewidth}{1pt} 
\renewcommand{\footrulewidth}{1pt}
\setlength{\arrayrulewidth}{1pt}
\setlength{\columnsep}{6.5mm}
\setlength\bibsep{1pt}

\twocolumn[
  \begin{@twocolumnfalse}
\noindent\LARGE{\textbf{Mesoscale phenomena in solutions of 3-methylpyridine, heavy water, and an antagonistic salt}}
\vspace{0.6cm}

\noindent\large{\textbf{Jan Leys,\textit{$^{a,b}$} Deepa Subramanian,\textit{$^{a,c}$} Eva Rodezno,\textit{$^{c}$} Boualem Hammouda,\textit{$^{d}$} and Mikhail A. Anisimov\textit{$^{a,c,\dag}$}}}\vspace{0.5cm}


\noindent \textbf{\small{Published in Soft Matter, \url{http://dx.doi.org/10.1039/c3sm51662h}}}
\vspace{0.6cm}

\noindent \normalsize{

We have investigated controversial issues regarding the mesoscale behavior of 3-methylpyridine (3MP), heavy water, and sodium tetraphenylborate (\ce{NaBPh4}) solutions by combining results obtained from dynamic light scattering (DLS) and small-angle neutron scattering (SANS). We have addressed three questions: (i) What is the origin of the mesoscale inhomogeneities (order of 100~nm in size) manifested by the ``slow mode'' in DLS? (ii) Is the periodic structure observed from SANS an inherent property of this system? (iii) What is the universality class of critical behavior in this system? Our results confirm that the ``slow mode'' observed from DLS experiments corresponds to long-lived, highly stable mesoscale droplets (order of 100~nm in size), which occur only when the solute (3MP) is contaminated by hydrophobic impurities. SANS data confirm the presence of a periodic structure with a periodicity of about 10~nm. This periodic structure  cannot be eliminated by nanopore filtration and thus is indeed an inherent solution property.  The critical behavior of this system, in the range of concentration and temperatures investigated by DLS experiments, indicates that the criticality belongs to the universality class of the 3-dimensional Ising model.}

\vspace{0.5cm}
 \end{@twocolumnfalse}

]

\footnotetext{\textit{$^{a}$~Institute for Physical Science and Technology, University of Maryland, 
College Park, MD 20742}}
\footnotetext{\textit{$^{b}$~Laboratorium voor Akoestiek en Thermische Fysica, Departement Natuurkunde en Sterrenkunde, KU Leuven, 3001 Leuven, Belgium.}}
\footnotetext{\textit{$^{c}$~Department of Chemical and Biomolecular Engineering, University of Maryland, College Park, MD 20742.}}
\footnotetext{\textit{$^{d}$~NIST Center for Neutron Research, National Institute of Standards and Technology, Gaithersburg, MD 20899}}
\footnotetext{\textit{\dag~Corresponding author. Tel: +1-301-405-8049; E-mail: anisimov@umd.edu} }

\section{Introduction}

Aqueous solutions of 3-methylpyridine (3MP) have received considerable attention in different contexts of soft matter research.  Part of this interest stems from the observation that 3MP is completely soluble in water \citep{andon1952}, but undergoes a closed-loop phase separation in heavy water\citep{cox1952} (\ce{D2O}).  As a consequence, pseudo-binary solutions of 3MP, water, and heavy water show a decreasing width of the phase-separated region with increasing normal to heavy water ratio \citep{narayanan1994}.  Systematic studies in these ternary solutions were used to verify theoretical predictions of the critical behavior, including effects such as critical exponent doubling and Fisher renormalisation \citep{sorensen1985,prafulla1992,prafulla1992b,prafulla1993,narayanan1994}.

An alternative mechanism to induce phase separation in solutions of 3MP and water is by the addition of salts \citep{prafulla1992b,narayanan1994,narayanan1995}. Studies on solutions of 3MP, water, and  sodium bromide (NaBr), near the critical point, showed the usual critical concentration fluctuations and the occasional presence of mesoscale inhomogeneities (order of 100 nm in size) \citep{jacob1998,anisimov2000a}. The mesoscale inhomogeneities were interpreted as due to a crossover phenomena between Ising and mean-field multi-critical behavior\citep{jacob1998,anisimov2000a}. However, this interpretation was challenged by further investigations \citep{hernandez2003,wagner2002,wagner2003,wagner2004,vanroie2003}. The results demonstrated that the inhomogeneities were essentially non-equilibrium (disappearing within 6 to 8~hours) and the normal 3-dimensional Ising critical behavior was recovered after the samples were fully equilibrated \citep{kostko2004}.  

Similar mesoscale inhomogeneities have also been observed in solutions of 3MP and water, far away from the critical point \citep{kostko2004,subramanian2011a}. These mesoscale inhomogeneities, unlike the ones mentioned above, seemed to remain stable over long periods of time (over a year or longer) \citep{subramanian2011a,subramanian2012a}. Detailed experiments on aqueous solutions of 3MP, and other small amphiphiles such as tertiary butyl alcohol, 2-butoxyethanol, and tetrahydrofuran, have revealed that these mesoscale inhomogeneities are Brownian, diffusive droplets, which occur only when the solute is contaminated by the presence of hydrophobic impurities \citep{subramanian2011a,subramanian2011b,li2011,habich2010,benderpecora2be}. The mesoscale droplets disappeared when the samples were heated, and reappeared upon cooling \citep{kostko2004,subramanian2011a}. The mesoscale droplets could be filtered out of the system and reinstated on the addition of trace amounts of a third, more hydrophobic, component \citep{subramanian2011b}. 

Solutions of 3MP and heavy water were extensively studied by small-angle neutron scattering (SANS) techniques upon the addition of different salts \citep{sadakane2006,sadakane2007a,sadakane2007b,sadakane2009,sadakane2011,sadakane2012}.  In these works, a number of different effects were reported:
(i) When a hydrophilic salt, such as LiCl (lithium chloride) or NaBr, was added mesoscale structures (order of 100 nm in size) were observed.  These structures were interpreted as either lamellar or spherical structures, depending on the nature and concentration of the salt \citep{sadakane2006,sadakane2007b}. (ii) When small amounts (5~mmol/L) of an amphiphilic salt (also known as ``antagonistic salt''), such as sodium tetraphenylborate (\ce{NaBPh4}) or tetraphenylphosphonium chloride (\ce{PPh4Cl}), were added to 3MP--\ce{D2O} solutions, a periodic structure was observed (with a typical periodicity of 10~nm) \citep{sadakane2007a,sadakane2009}.  This structure was interpreted in terms of the charge density wave model introduced by Onuki and Kitamura \citep{onuki2004} and was considered a consequence of the different affinities of the ions for the components in the solution \citep{sadakane2007a,sadakane2011}. (iii) When a higher concentration of \ce{NaBPh4} (85~mmol/L) was added to a solvent-rich 3MP--\ce{D2O} solution, a phase transition between a low temperature ordered ``onion-like'' phase and a high temperature disordered phase was observed \citep{sadakane2009}.  In the ordered phase, SANS data indicated the presence of a lamellar structure (with a typical periodicity of 17~nm), and confocal microscopic images showed the presence of microscopic droplets (order of tens of microns in diameter).  In the disordered phase, the charge density wave structure was observed. (iv) The critical behavior of the concentration fluctuations was studied in near-critical solutions  of 3MP, heavy water, and \ce{NaBPh4} by Sadakane \ea\citep{sadakane2011} The authors concluded that the effective confinement induced by the periodic structure changed the universality class of the critical behavior from 3-dimensional Ising criticality to 2-dimensional Ising criticality.

In this work, we bring these two research tracks together by a combined light and neutron scattering study of solutions of 3MP, heavy water, and \ce{NaBPh4}. We have addressed three questions: (i) What is the the origin of the mesoscale inhomogeneities (order of 100~nm in size) manifested by the slow mode in dynamic light scattering? (ii) Is the periodic structure observed in the presence of an antagonistic salt an inherent property of this system? (iii) What is the nature of critical behavior? Is it affected by the presence of the periodic structure? Finally, we present some preliminary results on the ``onion-like'' system. 

\section{Materials and methods\protect\footnote{Certain trade names and company products are identified in order to specify adequately the experimental procedure. In no case does such identification imply recommendation or endorsement by the National Institute of Standards and Technology, nor does it imply that the products are necessarily the best for the purpose.}}

\subsection{Materials}

In this work, the binary solution 3MP--\ce{D2O}, and the ternary solution 3MP--\ce{D2O}--\ce{NaBPh4} have been studied. The following chemicals have been used:
\begin{enumerate}
\item Heavy water, 99.851~mass\%, Atomic Energy of Canada Limited, Port Hawkesbury, N.S.
\item 3-Methylpyridine, 99.5+~\%, Aldrich.  Product number 23627-6, lot JS08507BQ.  Stated purity (CoA): 99.7~\%.
\item Sodium tetraphenylborate, ACS reagent $\geq$99.5~\%, Sigma-Aldrich.  Product number T25402, lot MKAA0457.  Stated purity (CoA): 99.54~\%.
\end{enumerate}

The concentrations of the samples studied in this work are presented in Table~\ref{table:samples}. The samples were prepared by cleaning the vial and drying them with dust-free nitrogen gas (99.9~\% purity purchased from Airgas).  The individual liquid components were filtered separately at room temperature by using 200~nm filters to remove dust particles (Nylon filters were used for heavy water, while Teflon filters were used for 3MP).  Components were then weighed and mixed at room temperature. If mesoscale droplets were observed in the final mixtures, then these mixtures were filtered at a low temperature (about 5~\textcelsius) by using a 20~nm Anopore filter (Anotop, Whatman), until the droplets disappeared.

\begin{table}
\small
\caption{Compositions of the samples reported in this paper}
\label{table:samples}
\centering
\begin{tabular}{cccccc}
\hline
& Sample & 3MP & 3MP & salt  \\
& & mass\% & mol\% & mmol/L \\
\hline
	& B1	& 27.43	& 7.52	   & 0		\\
	& B2	& 27.10	& 7.40	   & 0		\\
	& T1	& 27.52	& 7.55	   & 7.07	\\
	& T2	& 27.65	& 7.59	   & 6.88	\\
	& T3   & 27.63	& 7.59   & 7.12	\\
	& T4   & 27.47	& 7.52	   & 7.08\\
	& T5   & 31.63	& 9.07	   & 5.99	\\
	& O1	& 8.68	      & 2.06   & 85.7	\\
\hline
\end{tabular}\\
\end{table}

\subsection{Dynamic light scattering}

Dynamic light scattering (DLS) experiments were performed with the setup described in Ref.~\citenum{subramanian2011a}. The sample vial was immersed in a silicone oil bath, whose temperature could be controlled with an accuracy of $\pm 0.1$~K.  The  silicone oil, which has a refractive index close to that of vial's glass, was used to reduce the spurious scattering or refraction of the laser light. The scattered light from the sample, in the homodyne mode, was collected by a double photomultiplier tube, mounted on a goniometer. The dynamic auto-correlation function was obtained by using a Photocor correlator in its logarithmic scale (``multitau'') mode and with cross-correlation between the two photomultiplier channels.  

The time ($t$) dependent intensity auto-correlation function $g_2(t)$ was fitted to one or two exponentially decaying relaxation modes \citep{berne1976}:  
\begin{equation}
g_2(t)-1 = \left[A_1 \exp{\left(-\frac{t}{\tau_1}\right)} + A_2 \exp{\left(-\frac{t}{\tau_2}\right)}\right]^2 \ ,
\label{eq:correlation}
\end{equation}
where $A_i$ are the amplitudes and $\tau_i$ the decay (relaxation) times.  The decay times are related to the diffusion coefficient $D_i$ of the molecules or droplets by \citep{berne1976}:
\begin{equation}
\tau_i = \frac{1}{D_iq^2} \ ,
\label{eq:tau}
\end{equation}
where $q = 4\pi n/\lambda \sin(\theta/2)$ is the scattering vector, $n$ is the refractive index of the medium, $\lambda$ is the wavelength of the incident light (for a He-Ne laser $\lambda$ = 633 nm) and $\theta$ is the scattering angle.  For monodisperse, spherical, and noninteracting droplets, the Stokes-Einstein relation can be applied \citep{berne1976}:
\begin{equation}
D = \frac{\kB T}{6\pi \eta R_\mathrm{h}} \ ,
\label{eq:diffusion}
\end{equation}
where $k_\mathrm{B}$ is Boltzmann's constant, $T$  is temperature, $\eta$ is the shear viscosity of the medium, and $R_\mathrm{h}$ is the hydrodynamic radius. 

The refractive index $n$ was determined by an Abbe Refractometer (Fisher Scientific), with an accuracy of $\pm 0.0005$.  Measurements were only taken at room temperature, by assuming that  $n$ does not significantly change over the accessed temperature range.

The viscosity values of the samples at non-critical conditions was obtained indirectly from dynamic light scattering after doping the samples with latex particles of a known radius \citep{eulisssorensen}.  For viscosity values of samples at near-critical conditions, the light scattering intensity from the critical fluctuations was much higher than that of the latex particles. In such cases, the viscosity data were extrapolated by using the temperature dependence reported by Oleinikova, Bulavin, and Pipich for the binary 3MP--\ce{D2O} solution\citep{oleinikova1997}, and by assuming that the viscosity does not change over the accessed temperature range and on the addition of salt. 

\subsection{Small-angle neutron scattering} 

SANS experiments were performed with the NG3 SANS instrument at the NIST Center for Neutron Research. As in light scattering, the essential measurement length scale in SANS is the inverse of the wavenumber $q$, where $q = 4\pi /\lambda \sin(\theta/2)$, $\lambda = 0.6$~nm is the neutron wavelength, and $\theta$ is the scattering angle.  In the SANS experiments,  $q$ was varied from 0.03~nm$^{-1}$ to 4~nm$^{-1}$, corresponding to length scales from 1.5~nm to 200~nm.

\section{Results and discussion}

\subsection{Origin of mesoscale inhomogeneities}

\begin{figure}
\centering
\includegraphics[width=\columnwidth]{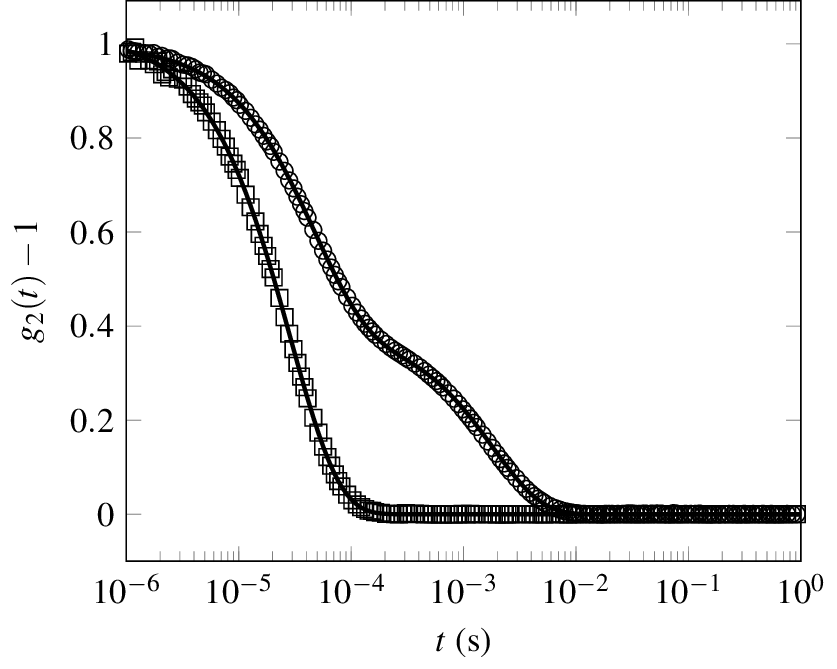}
\caption{\label{fig:corr_salt} Normalized intensity auto-correlation functions obtained from dynamic light scattering for a ternary 3MP--\ce{D2O}--\ce{NaBPh4} solution (sample T1) at $T=25$~\textcelsius\ and a scattering angle $\theta = 90$\textdegree. Circles represent the correlation function obtained initially, while squares represent the correlation function obtained after cold-filtering the sample with 20~nm filters. The black lines are fits to the form of Eq.~\ref{eq:correlation}.}
\end{figure}

Figure~\ref{fig:corr_salt} shows the intensity auto-correlation function observed from a solution of 3MP, heavy water, and \ce{NaBPh4} (sample T1) at a temperature of 25~\textcelsius\ and at a scattering angle of 90\textdegree. The correlation function shows the presence of two relaxation modes: a fast mode with a relaxation time of 56~$\mu$s and a slow mode with a relaxation time of 3.6~ms. The fast mode corresponds to molecular diffusion, with a diffusion coefficient of $4.7 \times 10^{-11}$~m$^2$/s. In accordance with Eq.~\ref{eq:diffusion}, this corresponds to an average hydrodynamic radius of 2~nm.  The hydrodynamic radius corresponds to the correlation length of concentration fluctuations and increases as the system approaches phase separation. 

The slower process, with a relaxation time of 3.6~ms, corresponds to the mesoscale droplets with an average hydrodynamic radius of 125~nm. In order to understand the origin of the slow mode, this ternary solution was filtered multiple times by using a 20~nm Anopore filter, at a low temperature (5~\textcelsius). The intensity auto-correlation function obtained after filtering the ternary solution at cold conditions is also shown in Fig.~\ref{fig:corr_salt}. The resultant correlation function shows no slow mode and thus no mesoscale inhomogeneities, but only the contribution from molecular diffusion. Similar mesoscale droplets were also observed in a binary solution of 3MP and heavy water (sample B1), which could be eliminated upon cold filtration; a similar phenomenon was also reported for binary solutions of 3MP and normal water \citep{subramanian2011a}. These results and the results obtained from other aqueous solutions containing small amphiphilic molecules studied in detail in Refs.~\citenum{sedlak2006a,sedlak2006b,sedlak2006c,subramanian2011a,subramanian2011b,subramanian2012a,li2011,benderpecora2be,habich2010}, led us to conclude that the slow mode originates only when such a system contains impurities (typically hydrophobic components). The presence of a salt does not trigger the slow mode. The slow mode can be eliminated by repeated filtrations under cold conditions and can be regenerated upon the addition of a controlled hydrophobic impurity, as was shown in Ref.~\citenum{subramanian2011b}. We also carried out experiments in the binary and ternary solutions by using another another source of 3MP. We noticed that the samples prepared from this source did not show the slow mode (mesoscale droplets), but only the fast mode (molecular diffusion), indicating that the previous 3MP source was contaminated by impurities which led to the mesoscale droplets.

\begin{figure}
\centering
\includegraphics[width=\columnwidth]{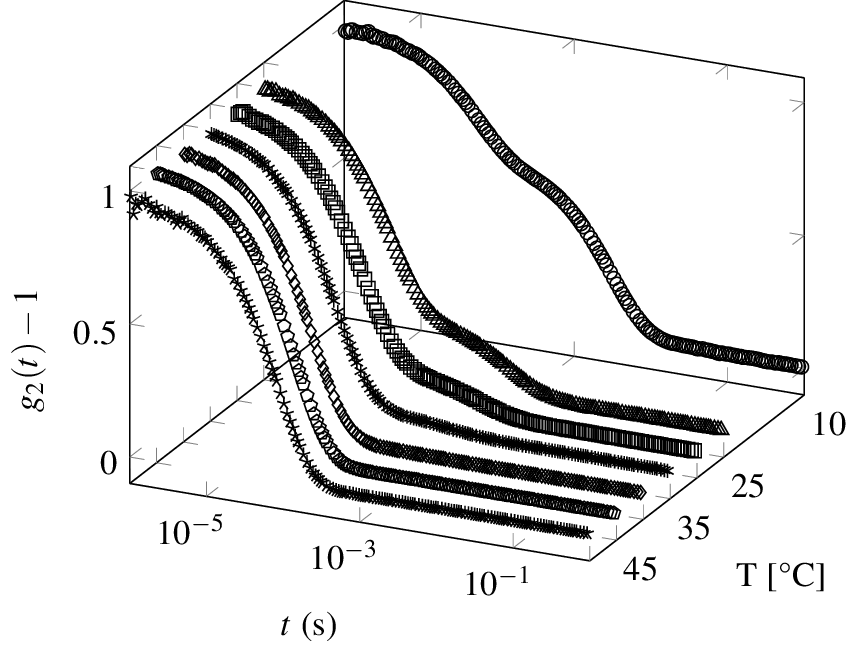}
\caption{\label{fig:corr_heat} Normalized intensity auto-correlation functions obtained from dynamic light scattering for a ternary 3MP--\ce{D2O}--\ce{NaBPh4} solution (sample T1) at a scattering angle $\theta = 90$\textdegree.  At low temperatures, two modes are observed: a slow mode and a fast mode, while at higher temperatures only the fast mode remains dominant.}
\end{figure}

Figure~\ref{fig:corr_heat} shows the intensity auto-correlation functions for the ternary system 3MP, heavy water, and \ce{NaBPh4} (sample T1) at various temperatures. This figure shows that as the temperature is raised, the slow mode corresponding to the mesoscale droplets disappears. This result is consistent with what has previously been observed in aqueous solutions of 3MP, tertiary butyl alcohol, and other small nonionic amphiphilic molecules \citep{sedlak2006a,sedlak2006b,sedlak2006c,subramanian2011a,subramanian2011b,subramanian2012a,li2011,benderpecora2be,habich2010}.  On the other hand, the fast mode corresponding to molecular diffusion, becomes more prominent as this system approaches phase separation.  The disappearance of the slow mode with increasing in temperature has been observed in several other systems, and in particular in aqueous tertiary butyl alcohol solutions \citep{subramanian2011a}, which does not show phase separation.  On the basis of this analogy, we can conclude that the slow mode genuinely disappears at high temperatures, and is not merely being obscured by the fast mode, which becomes dominant on approaching the phase separation.

\begin{figure}
\centering
\includegraphics[width=\columnwidth]{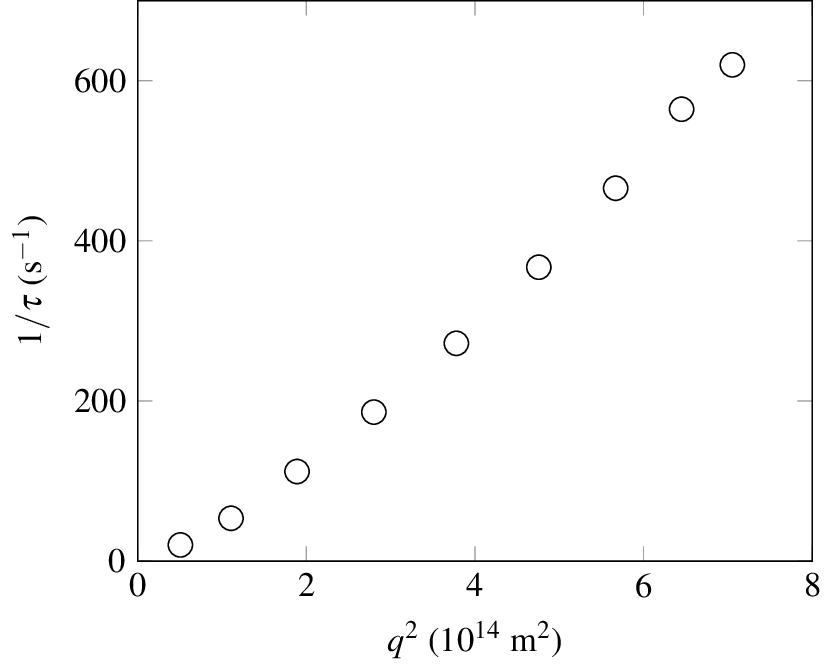}
\caption{\label{fig:diffusive_slow} Wave number dependence of the relaxation rate of the slow mode for a ternary system of 3MP--\ce{D2O}--\ce{NaBPh4} (sample T1) at $T=25$~\textcelsius.}
\end{figure}

\begin{figure}
\centering
\includegraphics[width=\columnwidth]{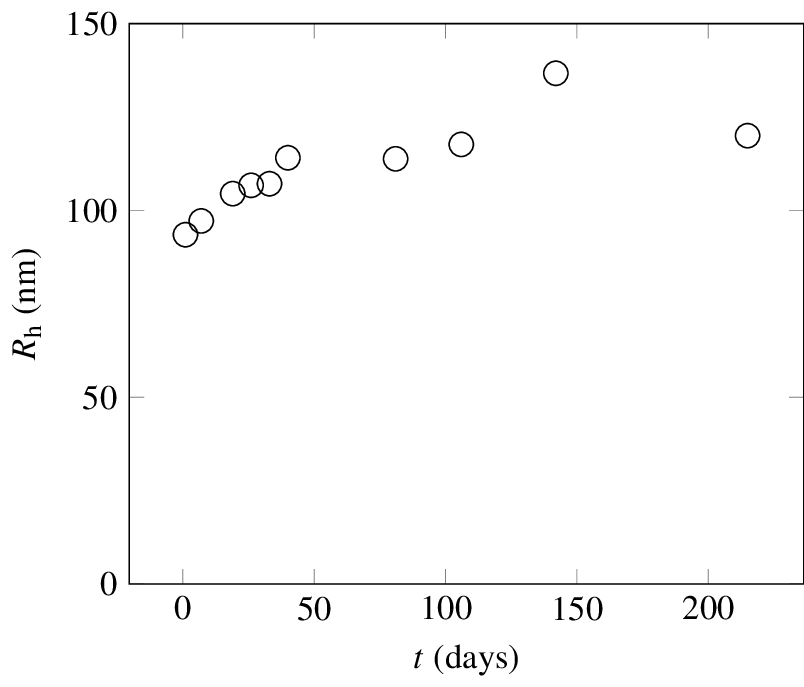}
\caption{\label{fig:time} Time dependence of the average hydrodynamic radius of the mesoscale droplets formed in a ternary 3MP--\ce{D2O}--\ce{NaBPh4} solution (sample T2) measured at $T=25$~\textcelsius\ and at scattering angle $\theta=90$\textdegree.}
\end{figure}

In order to understand the characteristics of the slow mode, the wavenumber dependence of the relaxation rate of the slow mode was determined. Figure~\ref{fig:diffusive_slow} shows the $q^2$ dependence of the slow mode observed in 3MP--\ce{D2O}--\ce{NaBPh4} solutions (sample T1) at a temperature of 25~\textcelsius. Some deviation from linearity in the wavenumber dependence of relaxation rate can be attributed to the polydispersity in the size of the mesoscale droplets. An unusual feature of these droplets is that they are stable over long periods of time. Figure~\ref{fig:time} shows the average size of the mesoscale droplets (sample T2) over a period of 8 months. It is seen from this figure that the size of the droplets remains almost unchanged over time, indicating that they could be in a kinetically arrested state. 

\subsection{\label{subsec:periodic}The periodic structure}

\begin{figure}[t]
\centering
\includegraphics[width=\columnwidth]{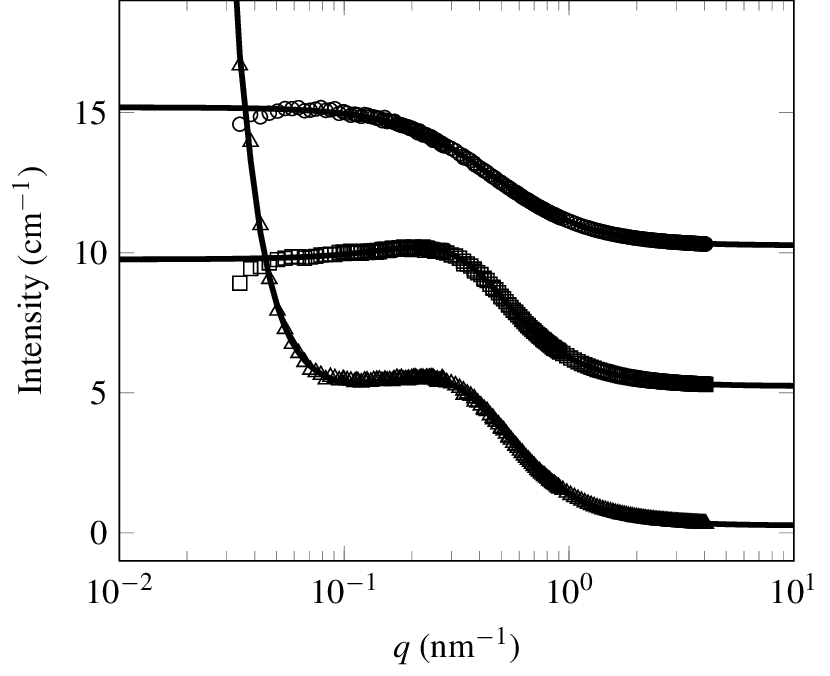}
\caption{\label{fig:SANS}SANS intensity data obtained at $T = 20$~\textcelsius. Triangles correspond to a ternary 3MP--\ce{D2O}--\ce{NaBPh4} solution (Sample T3).  Squares represent the same sample after filtering it with 20~nm filters at cold conditions. Circles represent a binary 3MP--\ce{D2O} solution after cold filtration (sample B1). The peak in the ternary system corresponds to a periodic structure (periodicity of $\sim 10$~nm) and the presence of a tail at low-$q$ corresponds to the mesoscale droplets. The black lines are fits to the form of Eq.~\ref{eq:OK} for the ternary system and Eq.~\ref{eq:OZ} for the binary system.  The curves have been shifted for display reasons; statistical error bars corresponding to one standard deviation are substantially smaller than the size of the symbols.}
\end{figure}

In this section, we report on the periodic structure observed in 3MP--\ce{D2O}--\ce{NaBPh4} solutions by carrying out small-angle neutron scattering experiments. Figure~\ref{fig:SANS} shows SANS intensity curves for a ternary solution of 3MP, heavy water, and \ce{NaBPh4} and a binary solution of 3MP and heavy water. The SANS data for the ternary system show the presence of a peak, which corresponds to a periodic structure. In addition, a tail at low wave number is observed, which corresponds to the mesoscale droplets described in the previous section. On filtering this solution under cold conditions the tail disappears, but the periodic structure remains unaffected. The resultant SANS intensity curve from the cold-filtered ternary solution is also shown in Fig.~\ref{fig:SANS}. The SANS intensity curve for the binary mixture 3MP--\ce{D2O} (after cold filtration) does not show the periodic structure or the mesoscale droplets, instead showing only the presence of concentration fluctuations. All these results lead us to the conclusion that the periodic structure is an inherent feature of 3MP--\ce{D2O}--\ce{NaBPh4} solutions, while the mesoscale droplets appear only when the solution is contaminated by the presence of impurities. 

The SANS data for the ternary solutions showing the presence of mesoscale droplets were analyzed by using the following equation:
\begin{equation}
I(q) =  \frac{A_3}{1+c_1q^2+c_2q^4}\ + \frac{A_4}{1+\left( 1- \frac{\kappa^2}{1+\lambda_\mathrm{D}^2q^2} \right)q^2\xi_\mathrm{SANS}^2} \,
\label{eq:OK}
\end{equation}
where $A_i$ are the amplitudes and $c_i$ are fitting parameters. 

The first part of the equation corresponds to the contribution from the mesoscale droplets. For a cold filtered solution, which does not show the presence of mesoscale droplets, the SANS intensity curves were analyzed by omitting this part of the equation. 

The second part of the equation corresponds to the periodic structure as suggested by the theory of Onuki and Kitamura \citep{onuki2004}. Here $\kappa$ is a coupling parameter which is associated with the different affinities of the ions for the different phases, $\lambda_\mathrm{D}$ is the Debye screening length, a measure for the effective range of the Coulombic interactions in this system, and $\xi_\mathrm{SANS}$ is the correlation length of concentration fluctuations. For $\kappa > 1$, the intensity exhibits a maximum, corresponding to a length scale
\begin{equation}
l_\mathrm{p} = 2\pi \frac{\lambda_\mathrm{D}}{\sqrt{\kappa-1}} \ ,
\label{eq:OK_length}
\end{equation}
which is interpreted as a repeat distance of the periodic structure. For $\kappa = 0$, the expression reduces to the usual Ornstein--Zernike expression, which we have used to analyze the data of the binary 3MP--\ce{D2O} water solution:
\begin{equation}
I_\mathrm{OZ}(q) = \frac{I_0}{1+q^2\xi_\mathrm{SANS}^2} \ .
\label{eq:OZ}
\end{equation}

\begin{figure}
\centering
\includegraphics[width=0.98\columnwidth]{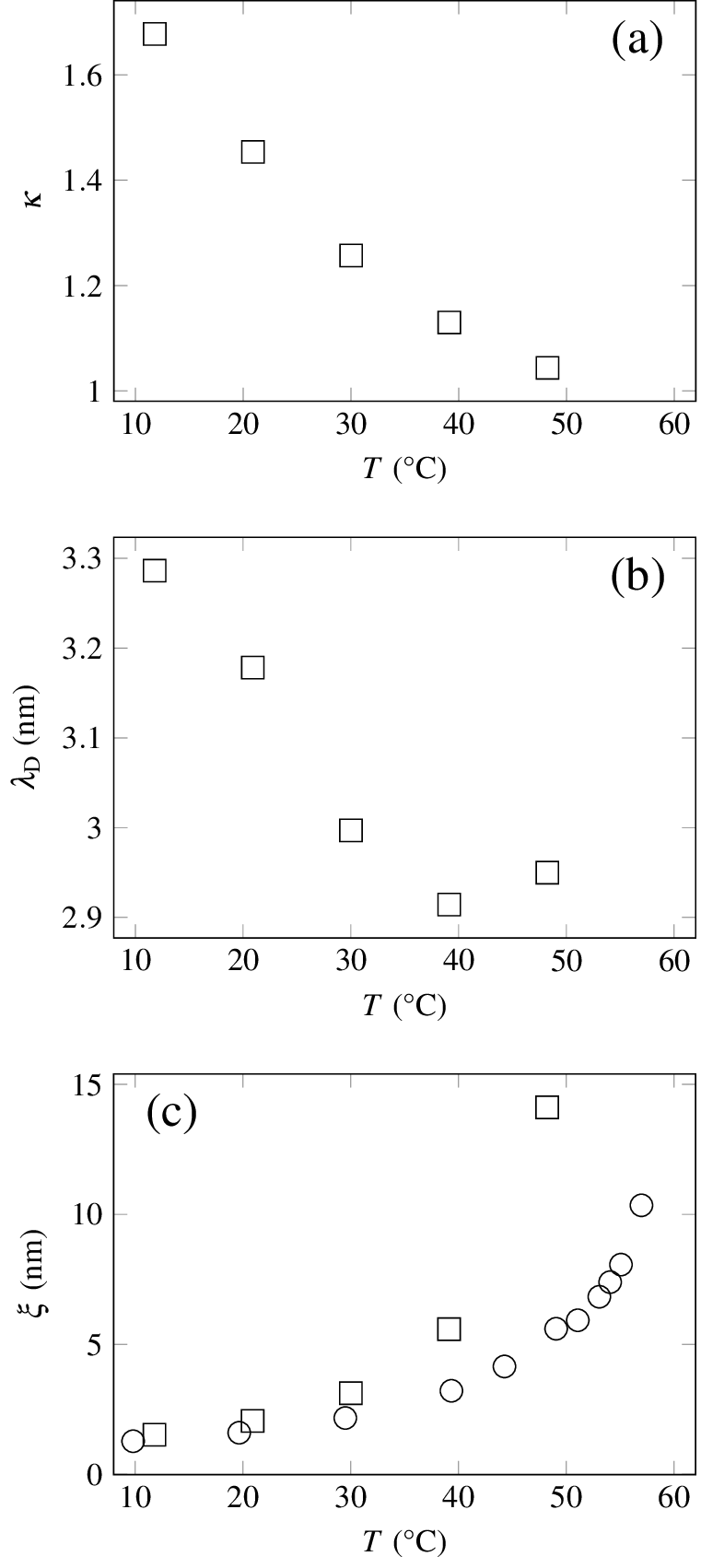}
\caption{\label{fig:SANS_fits}Parameters obtained from fitting Eq.~\ref{eq:OK} to the SANS intensity data. Data corresponds to a ternary 3MP--\ce{D2O}--\ce{NaBPh4} solution after cold-filtration (sample T3).
(a) Coupling parameter $\kappa$.
(b) Debye screening length $\lambda_\mathrm{D}$.
(c) Correlation length of concentration fluctuations $\xi_\mathrm{SANS}$. Circles correspond to the correlation length $\xi_\mathrm{DLS}$ obtained by analyzing DLS data. The composition of the sample studied by DLS (sample T4) is close to the composition of the sample studied by SANS (sample T3).
}
\end{figure}

Figure~\ref{fig:SANS_fits} shows the coupling parameter $\kappa$, the Debye screening length $\lambda_\mathrm{D}$, and the correlation length $\xi_\mathrm{SANS}$ for a 3MP--\ce{D2O}--\ce{NaBPh4} solution (sample T3) obtained by analyzing the SANS data. As seen from Fig.~\ref{fig:SANS_fits}(a), the decrease in the value of $\kappa$ upon approaching phase separation shows that the asymmetry in the affinities of the ions towards the different phase decreases. Fig.~\ref{fig:SANS_fits}(b) shows that the Debye screening length $\lambda_\mathrm{D}$ remains almost unchanged on increasing the temperature. An independent analytical value of $\lambda_\mathrm{D}$ can be determined by using the following expression \citep{jonesbook}: 

\begin{equation}
\lambda_\mathrm{D}^2 = \frac{\epsilon\epsilon_0k_\mathrm{B}T}{2c_0e^2} \ ,
\label{eq:debye}
\end{equation}
where $\epsilon$ is the dielectric permittivity of the solution, $\epsilon_0$ is the dielectric permittivity of vacuum, $k_\mathrm{B}$ is Boltzmann's constant, $T$  is temperature, $e$ is the ionic charge, and $c_0$ is the salt concentration. The estimated value of $\lambda_\mathrm{D}$ for our system (sample T3) at ambient conditions is 2.6~nm. This is close to the value obtained from analyzing the SANS data. 

Figure~\ref{fig:SANS_fits}(c) shows that the correlation length $\xi_\mathrm{SANS}$ increases as the temperature is raised, indicating approach to phase separation. Figure~\ref{fig:SANS_fits}(c) also shows another correlation length $\xi_\mathrm{DLS}$ which is obtained by analyzing DLS data for sample T4, in accordance to Eqs.~\ref{eq:correlation}--\ref{eq:diffusion}. Sample T4 has a concentration close to that of sample T3, which was studied by SANS. Figure~\ref{fig:SANS_fits}(c) indicates that the two correlation lengths thus obtained are quite different.

A depiction of the structure and formation of this periodic structure, consistent with our view as obtained form these experiments can be found in Figure 2(b) from Ref.~\citenum{sadakane2009}, while further details on the theory can be found in Refs.~\citenum{onuki2004} and \citenum{onuki2011}.

\begin{figure}
\includegraphics[width=\columnwidth]{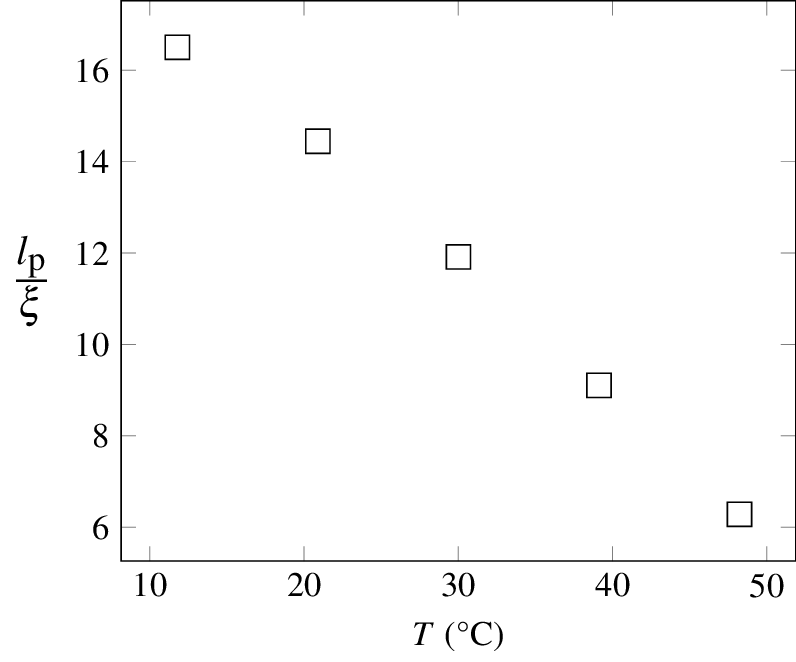}
\caption{\label{fig:ksi_lp_ratio} Ratio of the length scale of the periodic structure, $l_\mathrm{p}$ to the correlation length of concentration fluctuations, $\xi_\mathrm{SANS}$. Parameters are obtained from fitting Eq.~\ref{eq:OK} to the SANS intensity data. Data correspond to a ternary 3MP--\ce{D2O}--\ce{NaBPh4} solution after cold-filtration (sample T3).}
\end{figure}

Figure~\ref{fig:ksi_lp_ratio} shows the ratio of $l_\mathrm{p}/\xi_\mathrm{SANS}$ as a function of temperature, obtained from analyzing the SANS data. From this figure we see that $\xi_\mathrm{SANS}\ll l_\mathrm{p}$. This leads us to conclude that in the range of concentrations and temperatures studied, the confinement induced by the periodic structure will not significantly affect the behavior of the concentration fluctuations.

The periodic structure was not detected by dynamic light scattering. This is likely due to the fact that the relaxation time associated with the periodic structure is outside the range of what can be detected by dynamic light scattering (tens of microseconds to a few seconds). 

\subsection{\label{subsec:critical}Nature of criticality}

In this section, we investigate the nature of the critical behavior in a near-critical ternary solution of 3MP, heavy water, and \ce{NaBPh4} from DLS experiments, and compare our results to that obtained from SANS experiments carried out by Sadakane \ea\citep{sadakane2011} The ternary solution of 3MP, heavy water, and \ce{NaBPh4}, as well as the binary solution 3MP--\ce{D2O}, exhibit lower critical points. For the binary solution, the critical point is observed at a temperature of $\Tc \approx 38$~\textcelsius\  and $x_\mathrm{c} = 7.4$~mol\% \citep{cox1952}. For the ternary solution with 6~mmol/L \ce{NaBPh4}, the critical point is observed at a temperature of $\Tc \approx 42$~\textcelsius\ and $x_\mathrm{c} = 9.1$~mol\% (determined from the condition where the volumes of the two phases are equal). The asymptotic critical behavior of the correlation length of the concentration fluctuations is given as \citep{fisher1967,anisimov1991}:
\begin{equation}
\xi = \xi_0 \left( \frac{\Tc-T}{\Tc} \right)^{-\nu} = \xi_0 \epsilon^{-\nu} \ ,
\label{eq:crit_xi0}
\end{equation}
where $\xi_0$ is the critical amplitude, \Tc\ is the critical temperature, and $\nu$ is the critical exponent for the correlation length.  

In binary solutions, it is proven that the experimentally observed value of the critical exponent $\nu$, with high accuracy, follows the theoretical value of 0.63, predicted for the universality class of 3-dimensional Ising model \citep{anisimov1991}. In ternary solutions, certain phenomena could affect the observed value of the critical exponent: (i) the experimental path, along constant concentration of the third component, may lead to a change in the observed exponent, the effect known as Fisher renormalization \citep{fisher1968,anisimovthoen,anisimov1991}. We have checked that our experiments are in a temperature and concentration range where the effect of Fisher renormalization is negligible. (ii) Another effect that may become significant in systems that exhibit periodicity, is that the periodic structure may restrict the growth of the correlation length in one-dimension, making this criticality 2-dimensional. This effect will become pronounced when the length scale of the periodic structure $l_\mathrm{p}$ is comparable to the length scale of the concentration fluctuations. Such an interpretation of SANS experimental results has been suggested by Sadakane \ea\citep{sadakane2011}

\begin{figure}[t]
\centering
\includegraphics[width=\columnwidth]{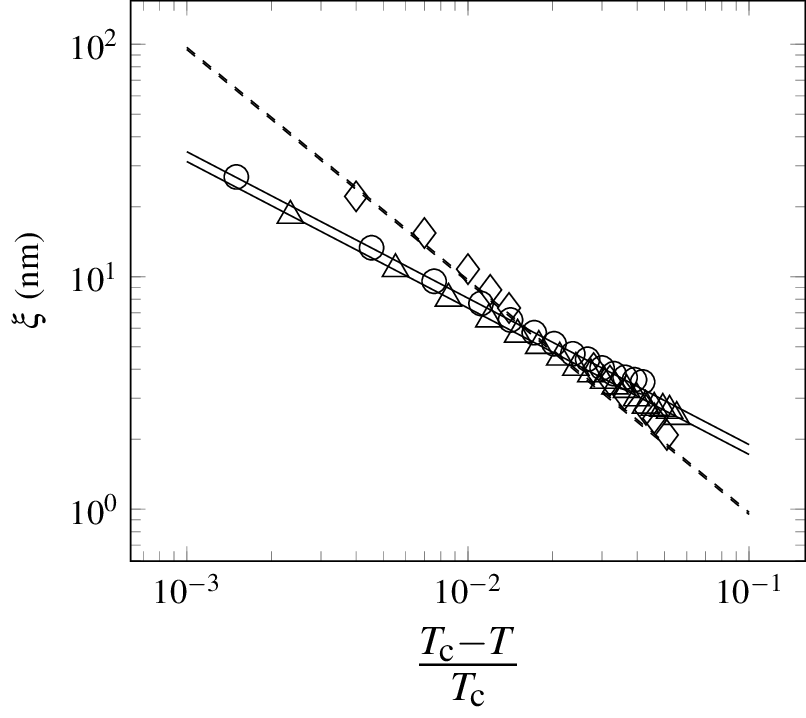}
\caption{\label{fig:crit} Correlation length versus reduced temperature for near-critical ternary 3MP--\ce{D2O}--\ce{NaBPh4} and binary 3MP--\ce{D2O} solutions. Circles (sample B2) and triangles (sample T5) correspond to correlation lengths $\xi_\mathrm{DLS}$ obtained from analyzing DLS experiments in accordance with Eqs.~\ref{eq:mc_ksi_gen}--\ref{eq:Db}. Diamonds correspond to correlation lengths $\xi_\mathrm{SANS}$ taken from Ref.~\citenum{sadakane2011}, which was obtained from the analysis of SANS experiments. The black solid lines are fits to the form of Eq.~\ref{eq:crit_xi0} with $\nu = 0.63$, while the dashed solid line is a fit to the form of Eq.~\ref{eq:crit_xi0} with $\nu = 1$.}
\end{figure}

\begin{table}
\small
\caption{Fit parameters for the critical behavior of the correlation length.  Parameters in square brackets have been kept fixed.}
\centering
\label{table:critical}
\begin{tabular}{ccccccc}
\hline
Sample		& \Tc		& $\xi_0$	& $\nu$	& $\chi^2_\nu$\\
			& \textcelsius	& nm		&		&\\
\hline
B2			& 37.645	& 0.44	& [0.630]	& 1.1	\\
			& 38.316  & 0.10	& [1]		& 189	 \\
			& 37.637	& 0.45	& 0.628	& 1.4	\\
		
T5			& 41.862	& 0.40	& [0.630]	& 1.8	\\
			& 42.780	& 0.09	& [1]		& 48\\
			& 41.947	& 0.35	& 0.667	& 0.9	\\				
\hline
\end{tabular}
\end{table}

For the analysis of the critical behavior, obtained from DLS experiments, an approach (explained in more detail in Refs.~\citenum{burstyn1983,jacob2001,kostko2007}) for the calculation of $\xi_\mathrm{DLS}$ from the diffusion coefficient $D$ is used. The diffusion coefficient $D$ can be decomposed in two contributions: $D = D_\mathrm{c} + D_\mathrm{b}$, with the critical contribution $D_\mathrm{c}$ and the background diffusion $D_\mathrm{b}$.  The critical contribution, $D_\mathrm{c}$ is given as:

\begin{equation}
D_\mathrm{c} = \frac{R_\mathrm{D} \kB T}{6\pi \eta \xi_\mathrm{DLS}} K(q\xi_\mathrm{DLS}) \left[ 1 + \left( \frac{q\xi_\mathrm{DLS}}{2} \right)^2 \right]^{z_\eta/2} \ ,
\label{eq:mc_ksi_gen}
\end{equation}
where $R_\mathrm{D}$ is a universal dynamic amplitude ratio (taken to be 1.05) \citep{sengersdas}, $K(q\xi_\mathrm{DLS}) \equiv K(x) = [3/(4x^2)][1+x^2+(x^3 - x^{-1})\arctan{x}]$ is the Kawasaki function, and where the factor $[1 + ( x/2 )^2]^{z_\eta/2}$ accounts for the divergence of the viscosity $\eta$ according to $\eta = \eta_\mathrm{b}(Q_0\xi)^{z_\eta}$.  Here $\eta_\mathrm{b}$ is the background viscosity, $Q_0$ a system-dependent amplitude and $z_\eta = 0.068$. The background contribution is given by
\begin{equation}
D_\mathrm{b} = \frac{\kB T}{16\eta_b\xi_\mathrm{DLS}} \left[ \frac{1+q^2\xi_\mathrm{DLS}^2}{q_\mathrm{C}\xi_\mathrm{DLS}} \right] \ ,
\label{eq:Db}
\end{equation}
where the wave number $q_\mathrm{C}$ has been fixed as $q_\mathrm{C}^{-1} \approx \xi_0 \approx 0.4$~nm.

The resultant correlation length $\xi_\mathrm{DLS}$ for the critical binary and ternary mixtures, obtained from analyzing the DLS data, is shown in Fig.~\ref{fig:crit}. The results for the critical behavior are summarized in Table~\ref{table:critical}. Various fits were performed: the 3-dimensional Ising value $\nu=0.630$, the 2-dimensional Ising value $\nu=1$, and $\nu$ as a free parameter.  The quality of the fit was quantified by a reduced chi-squared value $\chi^2_\nu$.

From Table~\ref{table:critical} and Fig.~\ref{fig:crit}, one can see that the fits with exponent values close to the 3-dimensional Ising value $\nu=0.630$ give the best description of the light scattering data, both for the binary and ternary solutions. In addition, the fit with $\nu=1$ gives a poor description of the data, and leads to an unphysical \Tc\ value (namely, the fitted value of \Tc\ is substantially larger than the temperature of the first point in the two-phase region observed experimentally). Also superimposed on Fig.~\ref{fig:crit} is the correlation length $\xi_\mathrm{SANS}$ taken from Ref.~\citenum{sadakane2011}, which was obtained by analyzing SANS data in accordance to the theory of Onuki and Kitamura \citep{onuki2004}. This figure shows that the correlation lengths for the ternary system obtained from these two different experiments and analysis techniques are indeed different. The correlation lengths obtained from analyzing the DLS data follow 3-dimensional Ising criticality, while the correlation lengths obtained by analyzing the SANS data follow 2-dimensional Ising criticality.

A possible reason for the discrepancy in the critical behavior observed by DLS and SANS experiments could be due to the fact that the correlation length observed from the two techniques are quite different. The correlation length obtained from DLS is calculated from the diffusion coefficient $D$ and the relaxation time $\tau$.  The relaxation time $\tau$, corresponding to the critical concentration fluctuations, is unaffected by the presence of the periodic structure, whose time scale is inaccessible by DLS. In contrast to this, the correlation length obtained from SANS is coupled with the length scale of the periodic structure and the Debye screening length. This may lead to an ``effective'' correlation length detected from SANS, which is not the same as the correlation length of the critical concentration fluctuations obtained from analyzing DLS data.

\subsubsection*{Note~~}

Recently, the coexistence curve of a near-critical 3MP--\ce{D2O}--\ce{NaBPh4} solution has been investigated.  The authors report a value of the order parameter critical exponent $\beta$ fully consistent with 3-dimensional Ising criticality \citep{troncoso2013}.

\subsection{\label{subsec:onion}``Onion-like'' system}

In this part, we investigate phase transition in a non-critical solution of 3MP, heavy water, and \ce{NaBPh4} (sample O1). The experiments correspond to a concentration range where a phase transition was reported between a low temperature phase with lamellar structure and a high temperature phase without these structures, separated by a two-phase region \citep{sadakane2009}.

Our visual observations confirm this phase sequence.  At low temperature, the samples appear as inhomogeneous viscous white liquids. At a temperature of about 45~\textcelsius, the sample is in a two-phase region.  The higher density phase is the disordered phase, while the lower density phase is the ordered phase.  It was not possible to perform dynamic light scattering experiments in the ordered phase, because the turbidity of the sample led to multiple scattering. However, it was possible to perform light scattering experiments in the disordered phase. The intensity auto-correlation functions from the disordered phase showed only a single exponential decay at short time scales, corresponding to the normal concentration fluctuations.  There was no suggestion of any other contributions in the spectrum. The depolarization ratio of this phase was also checked by light scattering to determine if there is any anisotropy. Within the experimental accuracy, we found no depolarization of the scattered light, and conclude that the pretransitional fluctuations in the disordered phase are most likely isotropic in nature.

\section{Conclusions}

Mesoscale behavior in 3MP, heavy water, and \ce{NaBPh4} solutions has been studied by DLS and SANS. We have confirmed that the occasionally observed mesoscale inhomogeneities (size of 100~nm) occur only when the solution is contaminated by impurities (typically hydrophobic components). These inhomogeneities are highly stable, long-lived, diffusive droplets that disappear at high temperature and can be physically removed by nanopore filtration. Such droplets are also observed in binary 3MP--\ce{D2O} solutions. 

The addition of an antagonistic salt leads to the formation of a periodic structure, which can be detected from SANS experiments. The periodicity is an inherent property of the solution and can be characterized by the charge-density wave model developed by Onuki and Kitamura \citep{onuki2004}. However, the correlation length of concentration fluctuations ($\xi_\mathrm{SANS}$) obtained from analyzing the SANS data by this model is not the same as the correlation length of concentration fluctuations ($\xi_\mathrm{DLS}$) obtained from analyzing DLS data. This poses a question: Do the correlation lengths, obtained by the two methods, have a different physical basis or is the discrepancy in the behavior of the two correlation lengths an artifact of the fitting technique? Is it possible that while $\xi_\mathrm{DLS}$ corresponds to the concentration fluctuations, $\xi_\mathrm{SANS}$ is an ``effective'' correlation length, which is somehow coupled with the other length scales present in the system, namely the Debye screening length $\lambda_\mathrm{D}$ and the periodicity $l_\mathrm{p}$? Moreover, the critical behavior of $\xi_\mathrm{DLS}$ follows 3-dimensional Ising criticality, while $\xi_\mathrm{SANS}$ follows 2-dimensional Ising criticality according to Ref.~\citenum{sadakane2011}. Further investigations are required to determine the exact reason for the difference in the behavior of the correlation lengths obtained from SANS and DLS in systems exhibiting a periodic structure.

\section*{Acknowledgements}

The authors thank David Weglein and Seong Kim for contributing to the light scattering experimental work. J.L. thanks Claudio A. Cerdeiri\~na, Patricia Losada--P\'erez and Jan Thoen for stimulating discussions.  D.S. and M.A.A. acknowledge useful interactions with A. Onuki. The research at the University of Maryland, College Park is supported by the Division of Chemistry of the National Science Foundation (grant no.\ CHE-1012052).  This work is based upon activities supported in part by the National Science Foundation under Agreement No.\ DMR-0944772.  J.L. acknowledges the Research Foundation -- Flanders (FWO) for a travel grant.

\footnotesize{
\bibliography{mesoscale_salt} 
\bibliographystyle{rsc} 
\balance
}

\end{document}